\documentstyle[12pt,epsfig]{article}
\newcounter{appendixc}
\renewcommand{\appendix}[1] {
        \refstepcounter{appendixc}
        \setcounter{equation}{0}
        \renewcommand{\theequation}{\Alph{appendixc}.\arabic{equation}}
	}

\begin{document}

\parindent=0cm

\title{Transaction costs: a new point of view}
\author{
Roberto~Baviera}
\maketitle

{\small \it
\begin{center}
Dipartimento di Fisica, Universit\`a dell'Aquila,
 \& I.N.F.M., Unit\`a dell'Aquila \\
  I-67100 Coppito, L'Aquila, Italy. \\
  E-mail : {\rm baviera@axtnt1.phys.uniroma1.it} \\
\end{center}
}
\maketitle

\vspace*{0.21truein}
\begin {abstract}
We consider a new approach to portfolio selection
in presence of transaction costs which
allows to map the problem into one without costs.
The proposed approach 
connects all the quantities of interest to 
exit times and probabilities to reach barriers.
This leads to analytic results in the Wiener case  
and to directly measurable quantities on a historical dataset 
in real markets.
\end{abstract}

\section{Introduction}

Optimal allocation of wealth among portfolios  
in presence of transaction costs
is a well known problem in 
literature~\cite{TakKlaAss,Constantinides,DumLuc,AkiMenSul,Serva}.
The investment we consider in this paper 
is on a risk-less bank account paying a fixed interest rate  
and on a risky asset with the logarithmic price increments
modeled by a Wiener process.

The general approach to face this problem 
is to follow the asset price during all the time and
sometimes ``control'' the portfolio buying or selling stocks.
This approach leads to problems of {\it singular} control Brownian motion
(see e.g.~\cite{Harrison,Karatzas} for a review)
which seem the most adequate in this context.
In an instantaneous control technique 
one has access (in principle) to the entire information;
i.e. the process is continuously observed, even
between two changes of trader's portfolio.
However, for all practical purposes,
the necessary information is limited only
to the moments 
the investor acts:
therefore, we shall focus our attention only on these moments
when something happens relevant from his point of view.
 
A trading rule is given if, each time the investor changes his portfolio,
one specifies:
\begin{itemize}
   \item which fraction of his capital has to be invested in the risky asset,

   \item when the investor should modify again his position;
\end{itemize}
then among all possible strategies 
the investor is interested in the optimal one.

Following Kelly~\cite{Kelly}
we consider an investor who desires to maximize
the growth rate of his capital.
It seems to be the most natural approach if
the trader is interested only in the long run behavior of his capital: 
in the following this kind of investor will be 
called speculator.

In financial markets assets have a bid and ask price.
In this case the asset value is no longer defined in a unique way:
it can take on any value in the bid-ask spread.
At every time one can buy assets at the ask price
or sell them obtaining the bid price,
which is always lower then the previous one.
In this paper we shall consider this spread in asset price as the only source
of trading costs.

The most common way to treat the portfolio optimization 
in presence of such costs~\cite{TakKlaAss,DumLuc} 
is to define an {\it absolute} asset value depending on the bid and ask price. 
However in a financial strategy it is natural to introduce 
a value {\it relative} to speculator's behavior,
i.e. an asset value which depends on the kind of operation 
(ask or bid) the market trader performs.
The value is a matter of conventions:
there are no financial reasons to prefer one choice or another.
Of course the optimal growth rate cannot depend on the convention used.
We shall show that this new point of view
allows a mapping of the portfolio optimization problem 
to a similar one without costs.
Therefore it connects a {\it singular} control problem
to information theory, 
with the main advantage that information  
{\it \`a la} Shannon~\cite{Shannon} of an asset price
can be measured on financial data and also 
computed in an elementary way for processes 
as the Wiener one considered here.

The paper is organized as follows:
in section {\bf 2} we state the portfolio selection problem.
We summarize the case of absence of costs
in a version appropriate for our purposes in section {\bf 3}. 
In section {\bf 4} the {\it relative} value 
approach is treated:
both the exact solution and an approximation of it are considered in detail
and we discuss 
the relations and differences between the {\it relative}
and {\it absolute} approach.
Finally section {\bf 5} is devoted to
summarize the results.
In appendix {\bf A} we deduce the  probabilities 
and average time to exit from a barrier in a Wiener process
and in appendix {\bf B} we discuss the {\it absolute} 
value approach in an approximate case.

\section{The model}

We consider a speculator who diversifies his portfolio in a risk-less 
bank account with a deterministic rate of growth $R$ and a risky
asset, for example a stock in a capital market. 

 The ask and bid prices of the stock's shares at 
calendar (continuous) time $t$ 
are related by the following equation  
\begin{equation}
   S^a_t = e^{\gamma} S^b_t\,\, ,
   \label{eq:price}
\end{equation}
where we consider the transaction cost $\gamma$ time independent.

The return at calendar time $t$ is defined 
\begin{equation}
   r_t \equiv \ln\frac{S^b_{t+1}}{S^b_t}\,\, .
\label{eq:return}
\end{equation}
 
The speculator is interested to choose a strategy which maximizes 
the exponential growth rate of his capital $W_t$
\begin{equation}
   \lambda \equiv \lim_{\Omega \to \infty} 
\frac{1}{\Omega} \ln\frac{W_{\Omega}}{W_0}\,\,.
   \label{eq:lyapunov}
\end{equation}

 In particular we shall consider the case 
where the trader modifies his position in the market only when 
a relevant change of the asset price appears.
This point of view is natural
in realistic situations where the trader  
waits until the return's variations are significant for him
and then rebalances his portfolio.
We call, as in~\cite{BavPasSerVerVul}, 
$\Delta$-trading time 
the rank of such an investment.

At $\Delta$-trading time $k$ the speculator keeps 
a fraction $l_k$ of his capital in assets,
while the remaining part is left in the bank account.
We shall consider only self-financing strategies
and we desire to repeat the game at every $\Delta$-trading time:  
therefore we shall limit our analysis to the fraction $l_k$
such that the capital is always strictly greater than zero.
The speculator waits until 
\begin{equation}
   r_{t,t_k} = \ln\frac{S^b_{t}}{S^b_{t_k}}\,\,
   \label{eq:return barrier}
\end{equation}
raises up to $\Delta^+_k$ or decreases to $-\Delta^-_k$
(with $\Delta^+_k, \Delta^-_k \geq 0$),
where both barriers are fixed  by the speculator at 
$\Delta$-trading time $k$.
When the return~(\ref{eq:return barrier}) ``hits'' one of the two barriers
the speculator 
can buy or sell assets.
We call exit time 
\[
{\cal T}_k \equiv t_{k+1}-t_k
\]
the lag between two portfolio changes.

To obtain a lighter notation 
we define the (final) {\it state} of the speculator $\eta$,
depending on the kind of operation (ask/bid) he will perform at
$\Delta$-trading time $k+1$:
\[
\eta = \left\{ 
	\begin {array} {cc}
	 a & (ask)\\
	 b & (bid)	
\end{array} \right. {\mathrm when \;\; the \;\; trader \;\;}
\begin {array} {l}
	 {\mathrm buys} \\
	 {\mathrm sells} \,\, .	
\end{array} 
\]

The fraction of the capital in assets is 
$
\exp(z_k \Delta^{z_k}_k) l_k W_k
$
just before the $k+1$-th trade and
$
l_{k+1} W_{k+1}
$
immediately after,
where the random variable $z_k$ can assume the values
\[	
	   z_k \equiv \left \{ {\begin {array} {ccl}
				  -1 & {\mathrm if} & r_{t_{k+1},t_k} < 0 \\
				  +1 & {\mathrm if} & r_{t_{k+1},t_k} > 0 \,\,.
			   \end{array} } \right . 
\]

In particular we shall consider the most natural investment
\begin{equation}
 \begin {array} {ccl}
	\eta = a & {\mathrm when} &   z_{k} = -1 \\
	\eta = b & {\mathrm when} &   z_{k} = +1 \,\,,
 \end{array}   \,\,\,\,
\label{eq:stato barriera}
\end{equation}
i.e. buy at the lower price and sell at the higher,
or equivalently the following condition must be satisfied
\begin{equation}	
 \begin {array} {ccc}
	\exp(-\Delta^{-}_k) l_k W_k & \leq & l_{k+1} W_{k+1} \\ 
	\exp(\Delta^{+}_k) l_k W_k & \geq & l_{k+1} W_{k+1}
 \end{array} \;\; {\mathrm if} \;\;
 \begin {array} {l}
	\eta = a \\
	\eta = b \,\, .
 \end{array}
\label{eq:l limits}
\end{equation}

In the proposed trading rule there are two classes of parameters,
connected to the two main characteristics of a financial strategy,
we have mentioned in the introduction:
\begin{itemize}
   \item $\{ l \}$ which specifies the fraction of the capital
	invested on assets,

   \item $\{ \Delta \}$ which is connected to the time the speculator 
	has to keep his position. 
\end{itemize}
We have split the optimal trading rule
in the two main questions for a speculator;
we shall show that it
will allow us to map the problem 
with transaction costs into one where no costs are present.

If we limit our attention to the trading times we can rewrite 
the growth rate of the capital
\begin{equation}
   \lambda = \frac{h}{\cal T} \,\, ,
   \label{eq:lyapunov discrete}
\end{equation}
where we call
\begin{equation}
   h \equiv \lim_{{\cal K} \to \infty} \frac{1}{\cal K} \sum^{\cal K}_{k=0} 
	\ln\frac{W_{k+1}}{W_{k}}\,\,
\label{eq:lyapunov definition}
\end{equation}
the mean lyapunov exponent of the capital,
${\cal K}$ is the number of investments (or equivalently exits of the process) 
up to the time $t$
and 
\begin{equation}
   {\cal T} \equiv 
  \lim_{{\cal K} \to \infty} \frac{1}{\cal K} \sum^{\cal K}_{k=1} 
	{\cal T}_{k}\,\,
   \label{eq:T definition}
\end{equation}
is the mean exit time.

In the case ergodic processes are considered, 
one can substitute the $\Delta$-trading time average in 
equations~(\ref{eq:lyapunov definition}) and (\ref{eq:T definition})
with the expectation value of the associate process. 

We notice that 
this analysis, where the returns are considered only
when they hit the barriers,
resembles Poincar\'e maps
in dynamical systems.
Going from continuous to discrete time setting,
the quantities of dynamical interest 
are simply rescaled by
the mean exit time ${\cal T}$~\cite{BadPol}.

The discrete time framework is clearly relevant 
in finance
because it allows to show the connection
of the capital growth rate, 
obtained via an optimal control policy, 
with quantities strictly related to
Shannon entropy~\cite{Shannon}. 
The main advantage is that 
these quantities can be directly measured on a historical dataset of 
financial assets~\cite{BavVerVul,BavVer} 
or computed via
elementary probability theory in simple cases. 
Furthermore, analyzing transaction costs 
from an unusual point of view this new framework 
will allow us to map the problem
into one without costs.

This approach is directly connected 
to the way a 
financial decision is taken 
in practice.
The speculator is interested 
in a growth rate which is optimal 
with respect to what
has happened up to $m$ $\Delta$-time steps before:
it is an intuitive fact that events far in the past
are irrelevant in the present portfolio selection.  
In other words, portfolio depends only on the last $m$ investments, i.e on 
a {\it finite} set of parameters $\{ l \}$ and  $\{ \Delta \}$,
where the Markovian order $m$ is chosen according to the
``memory'' of the asset process~\cite{BavVerVul}.
 
In this paper we shall consider the no-memory case 
modeling the returns with
\begin{equation}	
 dr_t = \mu dt + \sigma dw_t \,\, ,
\label{eq:wiener}
\end{equation}
where $\mu$ is the drift and $w_t$ a Wiener process with unitary variance.

We can safely assume a null risk-free interest rate $R$. 
The case $R > 0$
can always be recovered by simply replacing $\mu$ with $\mu - R$
and adding $R$ to the growth rate of the capital~(\ref{eq:lyapunov}).

\section{Absence of costs}

Using the notation introduced above
we summarize in this section
the well known case of portfolio selection in absence of transaction costs.

The capital at $\Delta$-trading time $k+1$ is given by 
\begin{equation}
W_{k+1} = \left[ 1 - l_k + l_k \exp(z_k \Delta^{z_k}_k) \right]  W_k  \,\, .
\label{eq:process no costs}
\end{equation}

We are considering a ``scale invariant'' trader:
maximizing the growth rate 
his behavior does not depend on the capital he has when he starts
an investment~\footnote{
The same property is true for all the utilities in the HARA class
(see e.g.~\cite{Merton} for a list
of the general characteristics of these functions).}.

In the case under consideration the return process has no memory
and no transaction costs are involved. 
The speculator repeats then
exactly the same game at every $\Delta$-trading time.
This implies that the optimal 
choice of both the fractions $\{ l \}$ and of barriers $\{ \Delta \}$  
do not depend on the considered time $k$.

The mean lyapunov exponent of the capital is 
\begin{equation}
 \begin {array} {l}
  h(l;\Delta^+,\Delta^-) = \\ 
  \;\;      p \ln \left[1+l(\exp(-\Delta^-)-1)\right] +
	(1-p) \ln \left[1+l(\exp(\Delta^+)-1)\right]  \,\, .
 \end{array} 
\label{eq:lyapunov no costs}
\end{equation} 
where $p=\pi(\Delta^+,\Delta^-)$ 
is the probability to exit from the lower barrier.
The capital growth rate 
as a function
of the chosen strategy, 
is obtained
dividing the mean lyapunov exponent~(\ref{eq:lyapunov no costs}) 
by the average exit time
${\cal T}=\tau(\Delta^+,\Delta^-)$.
We have derived in the appendix
the values of $\pi$   
(equation~(\ref{eq:pi})) and $\tau$ (equation~(\ref{eq:tau}))
for the Wiener process~(\ref{eq:wiener}).

We shall look for the maximum capital growth rate
finding first the optimal fraction $l$ at fixed barriers $\{ \Delta \}$
and then searching the optimal $\Delta$s.

The optimal $l$ satisfies 
\begin{equation}
   \l(\Delta^+,\Delta^-) = 
	\frac{1-p}{1-\exp(-\Delta^-)} -
	\frac{p}{\exp(\Delta^+)-1} \,\, .
   \label{eq:l Kelly}
\end{equation} 

Substituting the optimal value of $l$~(\ref{eq:l Kelly}) 
in equation~(\ref{eq:lyapunov no costs}) we can write 
the mean lyapunov exponent 
\begin{equation}
   h = p \ln \frac{p}{q} +  (1-p) \ln \frac{1-p}{1-q}  \,\,
   \label{eq:lyapunov Kelly}
\end{equation}
as the Kullback entropy~\cite{Kullback} between the probabilities
$p$ and $q$, 
where $q=q(\Delta^+,\Delta^-)$ is the 
martingale probability~(\ref{eq:martingale}).
As pointed out by Kelly~\cite{Kelly}
in a particular case,
relation~(\ref{eq:lyapunov Kelly}) plays a central role 
in log-optimal portfolio selection:
the growth rate obtained modifying the portfolio
according to an optimal control strategy~\cite{Harrison,Karatzas}
is linked to 
the entropy of the underlying process in information theory~\cite{Shannon}. 
Speculator's capital growth rate in absence 
and, as we shall show, in presence of transaction costs 
depends on the information present in the asset price.
Such an information  
can be measured in financial series~\cite{BavVerVul,BavVer} 
using a technique
inspired by the Kolmogorov $\epsilon$-entropy~\cite{Kolmogorov}.

\begin{figure}[htb]
 \begin{center}
  \resizebox{0.7\textwidth}{!}{\includegraphics{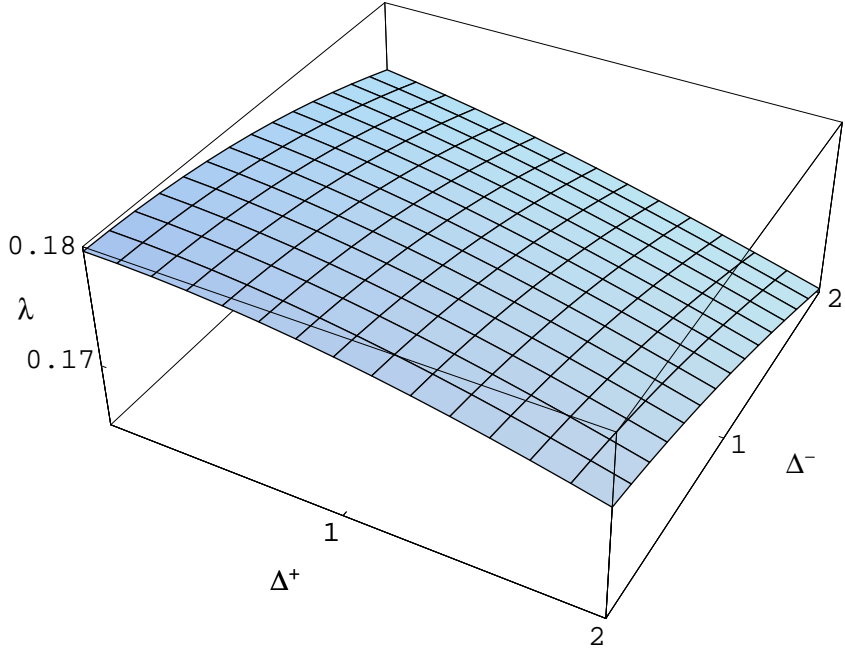}}
  \protect\caption{Capital growth rate in absence of costs 
for the optimal choice of $l$
as a function of $\Delta^+$ and $\Delta^-$.
The parameters chosen are $\mu=0.1$ and $\sigma=1$.}
  \label{fig:lyapunov no costs}
 \end{center}
\end{figure}

In Figure~\ref{fig:lyapunov no costs} we have plotted the 
capital growth rate $\lambda(\Delta^+,\Delta^-)$ for
the optimal choice of $l$. 
We notice that 
it is a non increasing function of its arguments.
The maximum is reached for $(\Delta^+,\Delta^-=0)$,
obtaining the well known result that the optimal policy is continuous.
It is an intuitive result due to the fact that a 
change in the portfolio does not cost anything 
and then the best solution is to use this {\it free} opportunity
to rebalance continuously the investment.

We also observe that the maximum $(\Delta^+,\Delta^-=0)$, 
is the only point
where the gradient of the growth rate is zero.
Therefore, if the speculator prefers to change his position
at finite $\Delta$s, 
the error in the growth rate is small, i.e. of the second order in $\Delta$
for small $\Delta$.

Performing the limit $\Delta^{+,-} \to 0$ 
in the equations~(\ref{eq:l Kelly},\ref{eq:lyapunov Kelly})
one obtains
the optimal capital fraction 
\begin{equation}
   l^* = \frac{\mu}{\sigma^2} + \frac{1}{2} \,\,
   \label{eq:l Kelly optimal}
\end{equation}
and the optimal growth rate  
\begin{equation}
   \lambda^* = \frac{\sigma^2}{2} \,\,\,\, {{l}^{*}}^2 \,\, .
   \label{eq:lyapunov Kelly optimal}
\end{equation}

We shall not allow the trader to borrow money from a bank 
or short selling of stock.
An optimal portfolio suggests
to keep the same fraction for ever ($l^*$ does not depend on $k$)
and,  
because we are thinking to model a portfolio including shares as risky assets,
a never-ending borrowing or short selling position 
does not appear realistic.
The considered cases correspond to 
\begin{equation}
-1 \leq \frac{2 \mu}{\sigma^2} \leq 1 \,\, .
   \label{eq:condition mu}
\end{equation}
If this ratio is $1$ 
the best solution is to transfer all the money to the stock,
instead the opposite limit corresponds to have no money
in the risky asset.
In the following we shall only consider 
transaction costs in the case with drift $\mu$ and variance $\sigma$
which satisfy condition~(\ref{eq:condition mu}). 

\section{The relative value approach}
 
In the portfolio selection in absence of costs there
is a unique asset price.
It is natural to associate the asset value to this price.
The value is a way to quantify the amount of money 
the trader has in his risky assets:
he diversifies his portfolio according to it,
investing a fraction of his capital $l_k$ which
depends on the asset value at $\Delta$-trading time $k$.
If transaction costs are present
the asset value is instead a convention.
One obtains the bid price selling the asset and buys it at the ask price.
The value can be any price 
between these two.   

To define the asset value,
two points of view seem natural:

\begin{itemize}
   \item An absolute {\it value} $S_t$ 
	which is considered the ``true price'',
	different from the ask and bid prices the trader finds in the market
	\[
 	\begin {array} {rcl}
		S^a_t & = & e^{\gamma_a} S_t \\
		S^b_t & = & e^{-\gamma_b} S_t
 	\end{array}  \,\,\,\,
	\]
	where $\gamma = \gamma_a + \gamma_b$ is the transaction cost
	defined in~(\ref{eq:price}).

   \item A value {\it relative} to the speculator's last trade:
   	it is the bid price when he sells and the ask price when he buys. 
\end{itemize}

The point of view considered in the literature
( see e.g.~\cite{TakKlaAss,DumLuc} and~\cite{AkiMenSul} and the references
there in)
is the {\it absolute} value approach\footnote{
Generally $\gamma_a =\gamma_b$ or 
one of the two parameters is kept equal to zero.
}.
In this approach an
asset is bought (or sold) at a price different from the value
of an asset of the same kind already included in trader's portfolio.
The absolute difference 
between the value and the price gives the cost.
In the appendix {\bf B} we summarize this approach in 
a approximate situation.

We stress here that the {\it absolute} value is a pure convention.
A financial asset is well defined only when 
a trader buys (sells) it;
its price is the ask (bid) price.
It seems natural to associate the same ``wealth'' not only 
to the assets just traded but also 
to all the others present in his portfolio even before this transaction.
In this way at every trading it exists only one value for the speculator.
This sounds reasonable because there is no difference between the
assets owned and the ones just bought (sold).
We are considering a value {\it relative} to the {\it state} 
of the speculator, as it as been defined in section {\bf 2}.
 
A consequence of a {\it relative} value is that
one has four possibilities
depending on 
an the kind of investment (ask/bid) the speculator is performing 
at $\Delta$-trading time $k$ (initial {\it state} $\xi$) and 
he will perform at $k+1$ 
(final {\it state} $\eta$)  depending on which barrier
the return~(\ref{eq:return barrier}) will hit.  
To underline this fact we shall denote the barriers 
not only with the $\Delta$-trading time of the operation
but also with the initial and final {\it state}.

We remind that
the bid and ask price are linked by relation~(\ref{eq:price}) 
and that the trader buys at the lower price and sells at the higher 
(see relations~(\ref{eq:stato barriera})).
The evolution of the capital invested in assets between time $k$ and $k+1$
is then:
\begin{equation}
\left\{ \begin {array} {cclccc}
{\displaystyle \frac{S^a_{k+1}}{S^a_k} } 
	l_k W_k & = & \exp(-\Delta_{aa;k}) l_k W_k 
	& {\mathrm if} &  \xi = a & , \;\; \eta = a \\
{\displaystyle \frac{S^b_{k+1}}{S^a_k} }
	l_k W_k & = & \exp(\Delta_{ab;k}-\gamma) l_k W_k 
	& {\mathrm if} &  \xi = a & , \;\; \eta = b \\
{\displaystyle \frac{S^a_{k+1}}{S^b_k} }
	l_k W_k & = & \exp(-\Delta_{ba;k}+\gamma) l_k W_k 
	& {\mathrm if} &  \xi = b & , \;\; \eta = a \\
{\displaystyle \frac{S^b_{k+1}}{S^b_k} }
	l_k W_k & = & \exp(\Delta_{bb;k}) l_k W_k 
	& {\mathrm if} &  \xi = b & , \;\; \eta = b \\
\end{array} \right.
\label{eq:cases} 
\end{equation}
where $\Delta_{ab;k}, \Delta_{ba;k} \geq \gamma$, i.e. one should at least
wait a time long enough to pay the costs!

We are considering the simple case of a no-memory process.
The only piece of information 
the speculator has to remember of his past
is whether at the previous $\Delta$-time step he has bought or sold assets,
i.e. his initial {\it state} $\xi$.
So, the barriers $\Delta_{\xi \eta;k}$ 
can only depend on the initial and final {\it states}
and the fraction $l_k$ only on the initial {\it state},
i.e when the $k^{th}$ investment is decided.

To obtain a lighter notation we introduce
\[
\left\{ \begin {array} {ccc}
\Delta^-_a &\equiv& \Delta_{aa} \\
\Delta^+_a &\equiv& \Delta_{ab}-\gamma \\
\Delta^-_b &\equiv& \Delta_{ba}-\gamma \\ 
\Delta^+_b &\equiv& \Delta_{bb} \,\, .
\end{array} \right. 
\]

The capital at $\Delta$-trading time $k+1$ is 
\begin{equation}	
W_{k+1} =
  \left[ 1 + l_{\xi} (\exp(z_{k}\Delta^{z_{k}}_{\xi}) - 1) \right] W_k 
	 \equiv  \omega_{\xi \eta} W_k \,\,,
\label{eq:relative process}
\end{equation}
where $z_k$ has been defined in section {\bf 2} as the sign of the return
between $k$ and $k+1$.
 
The problem is then completely defined by a Markovian transition matrix  
between the initial and final {\it state} 
\begin{equation}	
V = \left( \begin {array} {cc}
	1-\alpha &  \alpha \\
	 \beta   &  1 - \beta 
 \end{array}  \right) \,\,,	
\end{equation}
where 	
\[
\left\{ \begin {array} {ccc}
	\alpha & = & 1 - \pi(\Delta^-_{a},\Delta^+_{a}+\gamma) \\
	\beta  & = & \pi(\Delta^-_{b}+\gamma,\Delta^+_{b}) \,\, .
 \end{array}  \right. 
\]

The probabilities of the {\it states} $a$ and $b$ satisfy the following relation
\[	
	\frac{p_a}{p_b}=\frac{V_{ba}}{V_{ab}}=\frac{\beta}{\alpha} \,\, .
\]

We have mapped the original portfolio selection 
into one where no costs are present but a Markovian
memory must be considered.
The capital growth rate for a speculator who follows this strategy is:
\begin{equation}
 \lambda = 
  \frac{\sum_{\xi,\eta} p_\xi V_{\xi \eta} \ln \omega_{\xi \eta}}
       {\sum_{\xi}p_\xi {\cal T}_\xi} \,\, ,
\label{eq:lyapunov relative}
\end{equation}
where we have defined an average exit time
which depends on the initial {\it state} $\xi$ of the trader:
\[
\begin {array} {ccl}
{\cal T}_a &=& \tau(\Delta^+_{a}+\gamma,\Delta^-_{a}) \\
{\cal T}_b &=& \tau(\Delta^+_{b},\Delta^-_{b}+\gamma) \,\, .
\end{array}
\]  

Let us comment
equation~(\ref{eq:lyapunov relative}), which is the core of the paper.
The selection of the optimal trading rule 
from the prospective of the {\it relative} value
has led us to reduce our task to an optimization problem with 
a stochastic return driven by a finite-state Markovian chain. 
The problem can be then faced with elementary probability theory
in discrete spaces!

In the following we shall find the 
solution for $\{l\}$ and $\{ \Delta \} $ which optimize the
growth rate~(\ref{eq:lyapunov relative}).
We consider then an ansatz, with $l$ and $\{ \Delta^+, \Delta^- \}$
independent from the initial {\it state}, 
which appears quite natural
in the case of absence of memory for the underlying process,
we are dealing with in this paper.
We discuss in detail this approximation underlining why it can be
relevant for practical purposes
and compare the results with the {\it absolute} value approach.

\subsection{Exact solution}
 
We have shown how the
portfolio selection in presence of
transaction costs 
has been mapped into a Markovian
problem in absence of costs.
We find here the optimal solution
following the same route of previous section
maximizing first $\{ l \} $ and then finding the optimal barriers.

The two optimal values of $l$ are
\begin{equation}
\begin {array} {ccl}
 l_a(\Delta_a^+,\Delta_a^-) &=& 
 {\displaystyle \frac{\alpha}{1-\exp(-\Delta_a^-)} -
  \frac{1-\alpha}{\exp(\Delta_a^+)-1}} \\
 l_b(\Delta_b^+,\Delta_b^-) &=& 
 {\displaystyle \frac{1-\beta}{1-\exp(-\Delta_b^-)} -
 \frac{\beta}{\exp(\Delta_b^+)-1}} \,\,.
 \end{array} 
\label{eq:l relative}
\end{equation}
Substituting $l_a$ and $l_b$ in~(\ref{eq:lyapunov relative}) one obtains
\begin{equation}
 \lambda = 
  \frac{\sum_{\xi,\eta} p_\xi V_{\xi \eta} 
	\ln {\displaystyle \frac{V_{\xi \eta}}{Q_{\xi \eta}} } }
       {\sum_{\xi}p_\xi {\cal T}_\xi} \,\, ,
\label{eq:lyapunov relative solution}
\end{equation}
where
\[	
Q = \left( \begin {array} {cc}
	1-q_a &  q_a \\
	 q_b   &  1 - q_b 
 \end{array}  \right) \,\,,	
\]
with
\[
\left\{ \begin {array} {ccl}
	q_a & = & 1 - q(\Delta^+_{a},\Delta^-_{a}) \\
	q_b  & = & q(\Delta^+_{b},\Delta^-_{b})  \,\, .
 \end{array}  \right.
\]
 
We notice that 
the numerator of equation~(\ref{eq:lyapunov relative solution})
is for a Markov chain the quantity equivalent to the
Kullback entropy of equation~(\ref{eq:lyapunov Kelly}),
obtained in the no memory case: 
this is the {\it available} information
introduced in a simpler case by~\cite{BavPasSerVerVul}.

Let us stress that, for fixed barriers $\{ \Delta \} $,
the optimal growth rate of speculator's capital
is linked to the Shannon entropy of the 
underlying process
and then in general to a measurable quantity
even if transaction costs are present.

The speculator tries then to select the barriers $\Delta$
that maximize the information 
per unit of time~(\ref{eq:lyapunov relative solution})
associated to the process.
We observe in equations~(\ref{eq:cases}) 
that the trader pays costs only when he changes his {\it state}:
every modification of his portfolio still remaining in the same {\it state}
costs nothing.
Of course he will use this {\it free} opportunity to rebalance 
his portfolio as often as he can, i.e
\begin{equation}
 \Delta^-_a,\Delta^+_b \to 0 \,\, .
\label{eq:Delta limite zero}
\end{equation}
 
In this case the growth rate of the capital becomes:
\begin{equation}
 \lambda(\Delta^+_a,\Delta^-_b) = \mu \,\,
  {\displaystyle\frac{\tilde{\beta} \left(
	\tilde{\alpha} \ln 
  {\displaystyle\frac{\tilde{\alpha}}{\tilde{q_a}}} +
	\tilde{q_a}-\tilde{\alpha} \right) +
	\tilde{\alpha} \left(
	\tilde{\beta} \ln 
  {\displaystyle\frac{\tilde{\beta}}{\tilde{q_b}}} +
	\tilde{q_b}-\tilde{\beta} \right)}
       {\tilde{\beta} \left(-1+(\Delta^+_a+\gamma)\tilde{\alpha}) \right) +
	\tilde{\alpha} \left(1-(\Delta^-_b+\gamma)\tilde{\beta}) \right)}} 
	\,\, ,
\label{eq:lyapunov relative zero}
\end{equation}
where
\[
\left\{ 
 \begin {array} {ccl}
  \tilde{\alpha} & = & {\displaystyle
   \frac{\partial \alpha}{\partial \Delta^{-}_a}|_{\Delta^{-}_a=0}}\\
  \tilde{q_a} & = & {\displaystyle 
   \frac{\partial q_a}{\partial \Delta^{-}_a}|_{\Delta^{-}_a=0}}
\end{array} \right. {\mathrm and} \,\,
\left\{	\begin {array} {ccl}
	 \tilde{\beta} & = & {\displaystyle
   \frac{\partial \beta}{\partial \Delta^{+}_b}|_{\Delta^{+}_b=0}}\\
	 \tilde{q_b} & = & {\displaystyle 
   \frac{\partial q_b}{\partial \Delta^{+}_b}|_{\Delta^{+}_b=0}} \,\, .
\end{array} \right.
\]
We notice that even in this case
the growth rate~(\ref{eq:lyapunov relative zero}) is a non increasing
function of its arguments.

The optimal fractions~(\ref{eq:l relative}) become 
\begin{equation}
 \begin {array} {ccl}
l_a (\Delta^{+}_a) & = & \tilde{\alpha} - \tilde{q_a}\\
l_b (\Delta^{-}_b) & = & \tilde{q_b} - \tilde{\beta} \,\,. 
\end{array}
\label{eq:l relative zero}
\end{equation}
\begin{figure}[htb]
 \begin{center}
  \resizebox{0.7\textwidth}{!}{\includegraphics{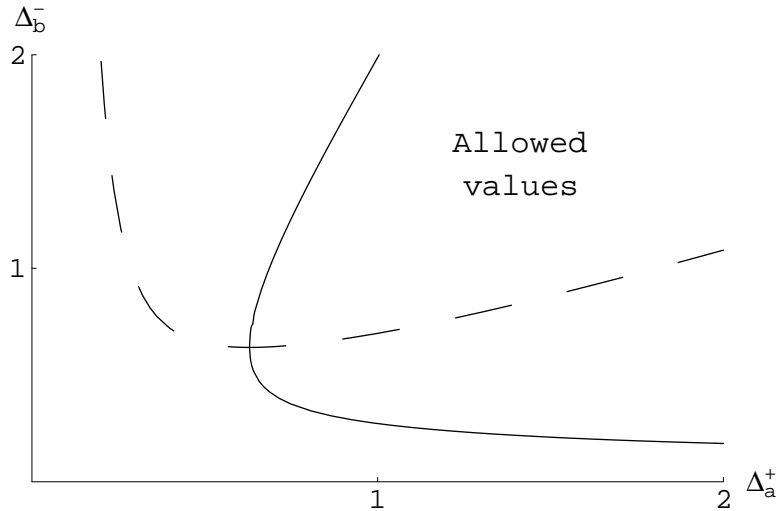}}
  \protect\caption{The values of $\Delta$s allowed by 
conditions~(\ref{eq:l limits III})
for $\mu / \sigma^2 =0.1$ and $\gamma=0.01$.
The full line represents $\nu^{-1}_a \left[ l_b(\Delta^-_b) \right]$
and the dashed 
$\nu^{-1}_b \left[ l_a(\Delta^+_a) \right]$.}
  \label{fig:l limits}
 \end{center}
\end{figure}

We remind that we are looking for a solution where the trader sells 
at the higher price and buys at the lower, and then we have to check that 
conditions~(\ref{eq:l limits}) are satisfied, which imply:
\begin{equation}
\left\{ 
\begin {array} {ccccl}
 \nu_a(\Delta^+_a) & \equiv & 
   {\displaystyle \frac{ e^{\Delta^+_a}\,l_a(\Delta^+_a)}
	{1+ (e^{\Delta^+_a}-1)\,l_a(\Delta^+_a) }}
	 & \geq &l_b(\Delta^-_b)\\ 
 \nu_b(\Delta^-_b) & \equiv & 
  {\displaystyle \frac{ e^{-\Delta^-_b}\, l_b(\Delta^-_b)}
	{1+ (e^{-\Delta^-_b}-1)\, l_b(\Delta^-_b)}}
	 & \leq &l_a(\Delta^+_a) \,\,.
\end{array} \right.
\label{eq:l limits II}
\end{equation}

It is easy to verify that $\nu_\xi$ are monotone and in particular
$\nu_a$ is a strictly increasing function of its argument and
 $\nu_b$ strictly decreasing.
Inverting~(\ref{eq:l limits II}) one obtains:
\begin{equation}
\left\{ 
	\begin {array} {ccl}
	 \Delta^+_a & \geq &\nu^{-1}_a \left[ l_b(\Delta^-_b) \right]\\
	 \Delta^-_b & \geq &\nu^{-1}_b \left[ l_a(\Delta^+_a) \right] \,\,.
\end{array} \right.
\label{eq:l limits III}
\end{equation}
 
In Figure~\ref{fig:l limits} we have shown the region  
of $\Delta$s allowed by condition~(\ref{eq:l limits III}).
We observe in particular that 
$\nu^{-1}_a \left[ l_b(\Delta^-_b) \right]$ 
and 
$\nu^{-1}_b \left[ l_a(\Delta^+_a) \right]$ 
cross at the point $\Delta^+_a = \Delta^-_b = \Delta$ 
where $\Delta$ is given by equation
\begin{equation}
 \frac{2 \mu}{\sigma^2} \sinh\left(\frac{\Delta}{2} \right) =  
 \sinh \frac{\mu}{\sigma^2} \left( \Delta +\gamma \right) \,\,.
\label{eq:soluzione barriera}
\end{equation}
We notice that it is possible to show after 
some algebra that 
the tangents of the two curves are parallel to the axes
in the crossing point 
$(\Delta^+_a , \Delta^-_b = \Delta)$.

\begin{figure} 
\begin{center}
  \resizebox{0.7\textwidth}{!}{\includegraphics{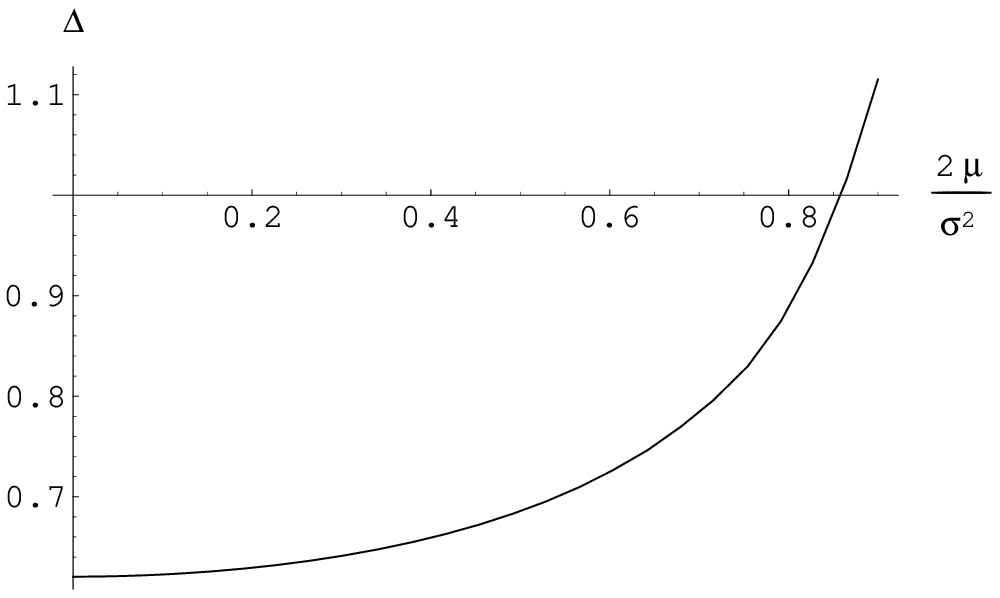}}
  \protect\caption{Optimal values of $\Delta$ {\it vs} $2 \mu/ \sigma^2$ for $\gamma=0.01$.}
  \label{fig:Delta relative exact}
 \end{center}
\vspace{1.2cm}
 \begin{center}
  \resizebox{0.7\textwidth}{!}{\includegraphics{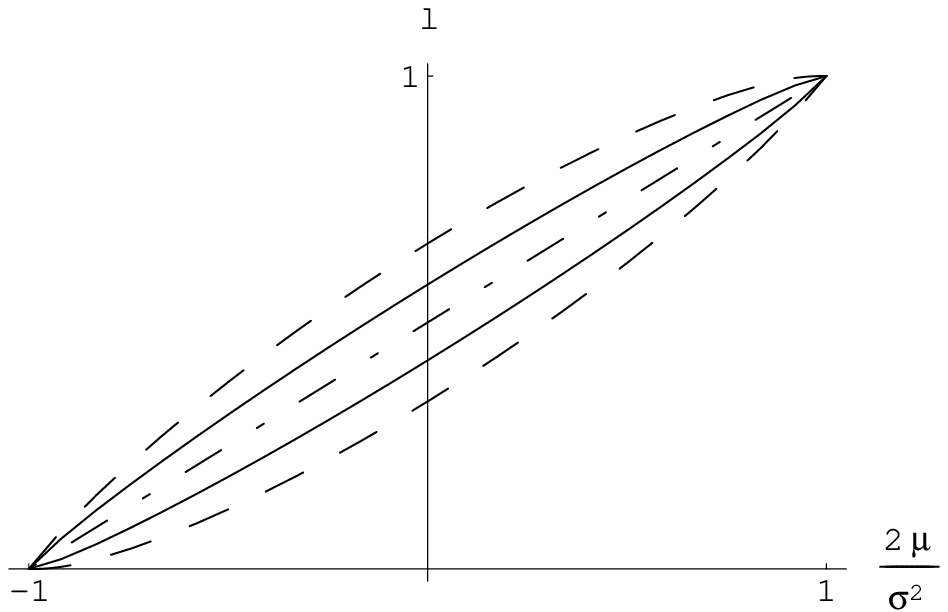}}
  \protect\caption{Optimal values of $l_a$ and $l_b$ {\it vs} 
  $2 \mu/ \sigma^2$. 
  The fraction of the capital $l_a$ is greater than 
  the value $l^*$ obtained in absence of costs (dot dashed line)
  while $l_b$ is lower.
  The dashed line corresponds to the value of $\gamma=0.1$
  and the full line to $\gamma=0.01$.}
  \label{fig:l relative exact}
 \end{center}
\end{figure}

Equation~(\ref{eq:soluzione barriera}) gives the optimal choice
for the barriers
of the portfolio selection problem in presence of transaction costs, 
because the growth rate~(\ref{eq:lyapunov relative zero})
is a non increasing function of the two barriers.
 
In Figure~\ref{fig:Delta relative exact}
we plot the optimal values of $\Delta$ as a function of $2 \mu/ \sigma^2$
obtained from equation~(\ref{eq:soluzione barriera})
for a particular choice of $\gamma$
observing that $\Delta$ goes to infinity when  $2 \mu/ \sigma^2$
approaches $1$.
We have considered only the positive values because
$\Delta$ is an even function of $\mu$.

One can also show that 
$(\Delta^-_a , \Delta^+_b =0;\Delta^+_a , \Delta^-_b = \Delta)$
is the only point where the gradient of 
the growth rate~(\ref{eq:lyapunov relative solution})
is the null vector.
This fact is particularly relevant for a speculator,
because the optimal solution, as a consequence 
of~(\ref{eq:Delta limite zero}), requires on average 
to rebalance the portfolio after a time equal to zero.
A null gradient of the capital growth rate
on the optimal solution implies, 
as in the absence of costs case,
that a suboptimal choice of the barriers causes a small
error in the capital growth rate.

In Figure~\ref{fig:l relative exact} we have shown the values 
of $l_\xi$~(\ref{eq:l relative zero})  on the optimal solution
for $\Delta$s~(\ref{eq:soluzione barriera}) for two different values 
of $\gamma$ and the fraction $l^*$ of the capital 
in the no cost case~(\ref{eq:l Kelly}).
After some algebra one can obtain that
\begin{equation}
\begin {array} {c}
l_a(\tilde{\Delta}) =  2 l^* - l_b(\tilde{\Delta}) \,\,, 
\end{array}
\label{eq:l relatione}
\end{equation}
i.e. $l^*$ is the mean value of 
$l_a(\tilde{\Delta})$ and $l_b(\tilde{\Delta})$ for all $\tilde{\Delta}$
and in particular on the optimal solution~(\ref{eq:soluzione barriera}).

The optimal growth rate of the portfolio in presence of transaction costs is
\begin{equation}
 \lambda_{O} = 
  \frac{\sigma^2}{2} l_a(\Delta) l_b(\Delta) \leq \lambda^* \,\, ,
\label{eq:lyapunov solution relative}
\end{equation}
where $\lambda^*$ is the one obtained in the no-costs 
case~(\ref{eq:lyapunov Kelly optimal}).
We notice that the optimal growth rate of the capital is 
proportional to
the square of the geometric average of $l_a$ and $l_b$
and $\lambda^*$ to the square of their arithmetic average.
The geometric average is always lower or equal to the arithmetic average
$l^*$ and equal only when $l_a=l_b$, i.e. for $2\mu/\sigma^2=-1,1$.
This fact has a simple interpretation: 
these are the cases where the trader maintains his position for ever;
in these situations no costs are payed, except at least once
when the investment begins.

\subsection{An approximate solution}

In this subsection we consider 
the simplest ansatz for speculator's strategy:
it will allow us to compare the two approaches of the {\it relative}
and {\it absolute} value and to suggest a feasible approximation
of the optimal trading rule.
We choose both the barriers and the fraction independent
on the initial {\it state} $\xi$.
This should be a reasonable approximation,
as explained in the no-costs case,
because we are dealing with a no memory process
and a growth rate maximizer.
We notice that conditions~(\ref{eq:l limits}) are automatically
satisfied once we have chosen a time independent value of $l$. 

\begin{figure}[htb]
 \begin{center}
  \resizebox{0.7\textwidth}{!}{\includegraphics{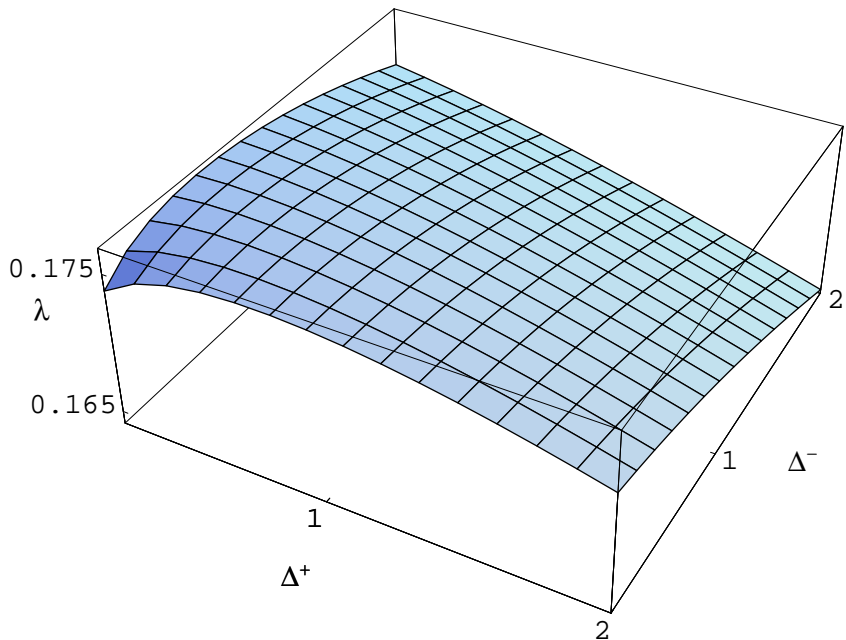}}
  \protect\caption{Capital growth rate for the optimal choice of $l$
as a function of $\Delta^+$ and $\Delta^-$.
The parameters are $\mu=0.1$, $\sigma=1$, $\gamma=0.01$.}
  \label{fig:lyapunov absolute costs}
 \end{center}
\end{figure}
 
The capital growth rate~(\ref{eq:lyapunov relative}) becomes 
\begin{equation}
 \lambda = 
  \frac{ p \ln \left[1+l(\exp(-\Delta^-)-1)\right] + 
     (1-p) \ln \left[1+l(\exp(\Delta^+)-1)\right]}
       {p {\cal T}_a + (1-p) {\cal T}_b} \,\, ,
\label{eq:lyapunov approx relative}
\end{equation}
where $p=\beta/(\alpha + \beta)$.
\begin{figure}
 \begin{center}
  \resizebox{0.7\textwidth}{!}{\includegraphics{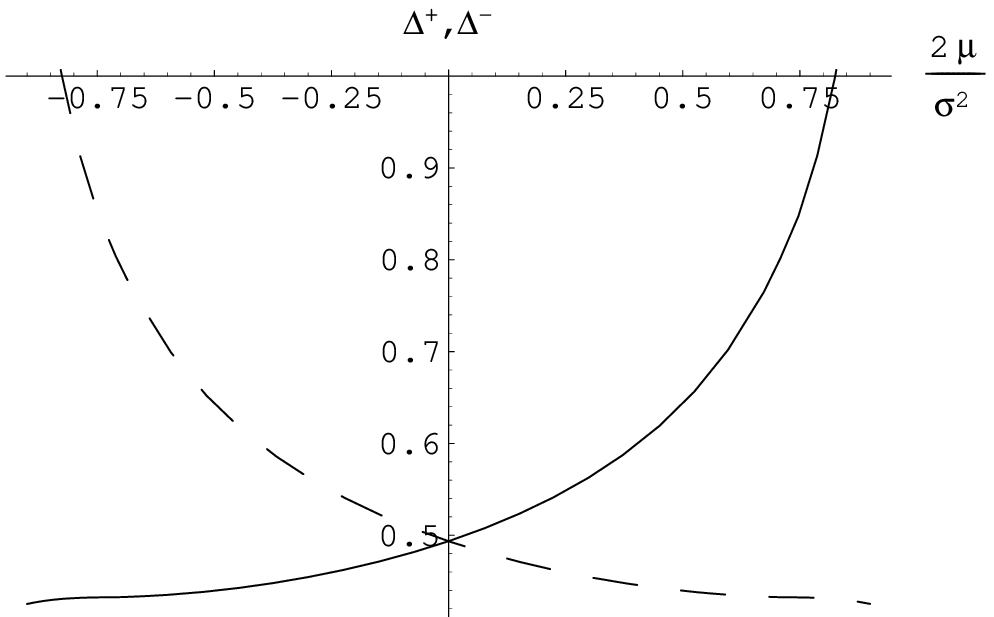}}
  \protect\caption{Optimal values of $\Delta^-$ (full line) and $\Delta^+$ (dashed line) {\it vs} $2 \mu/ \sigma^2$ for $\gamma=0.01$.}
  \label{fig:delta relative costs}
\end{center}
\vspace{1.2cm}
\begin{center}
  \resizebox{0.7\textwidth}{!}{\includegraphics{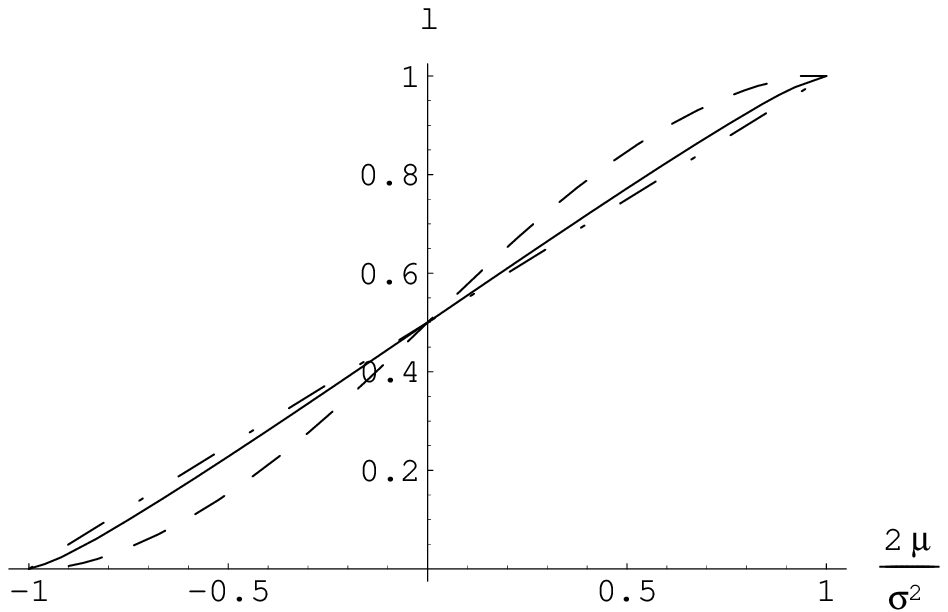}}
  \protect\caption{Optimal values of $l$ {\it vs} 
  $2 \mu/ \sigma^2$. The dashed line corresponds to the value of $\gamma=0.1$
  and the full line to $\gamma=0.01$. 
  We have plotted also the value of $l$ in absence of costs (dot dashed line).}
  \label{fig:l absolute costs}
 \end{center}
\end{figure}

We have mapped the problem into one in absence of costs
considered in section {\bf 3}
with probability $p$ to exit from the lower barrier
and average exit time ${\cal T}= p {\cal T}_a + (1-p) {\cal T}_b$.
The optimal value of $l$ is given by equation~(\ref{eq:l Kelly}).
Substituting this value in equation~(\ref{eq:lyapunov approx relative}) one 
obtains again that the growth rate can be written as 
the ratio between the Kullback entropy~(\ref{eq:lyapunov Kelly})
and the average exit time ${\cal T}$.

In Figure~\ref{fig:lyapunov absolute costs} we have plotted
the capital growth rate~(\ref{eq:lyapunov discrete}) 
as a function of the barriers
$\Delta^+$ and $\Delta^-$ 
calculated for the optimal $l$~(\ref{eq:l Kelly}).
We observe that the maximum is reached 
for finite values of the barriers.
Thus  
the speculator (on average) changes his portfolio 
after a finite time,
i.e. he follows a discrete time
trading rule.

In Figure~\ref{fig:delta relative costs} we plot 
the values of the barriers for which the optimal growth rate
is reached.
The fraction $l$ of the capital computed
on the optimal barriers has been plotted 
in Figure~\ref{fig:l absolute costs}.
We notice that 
\begin{eqnarray*} 
\Delta^-(-\mu) & = & \Delta^+(\mu) \\
\Delta^+(-\mu) & = & \Delta^-(\mu) \\
l(-\mu)         & = & 1-l(\mu)
\end{eqnarray*}
as a consequence of the symmetry 
\begin{equation}
\lambda_{-\mu}(l;\Delta^+,\Delta^-)=\lambda_{\mu}(1-l;\Delta^-,\Delta^+)-\mu 
\label{eq:simmetry}
\end{equation}
of the capital growth rate~(\ref{eq:lyapunov approx relative}).  

\begin{figure}[htb]
 \begin{center}
  \resizebox{0.7\textwidth}{!}{\includegraphics{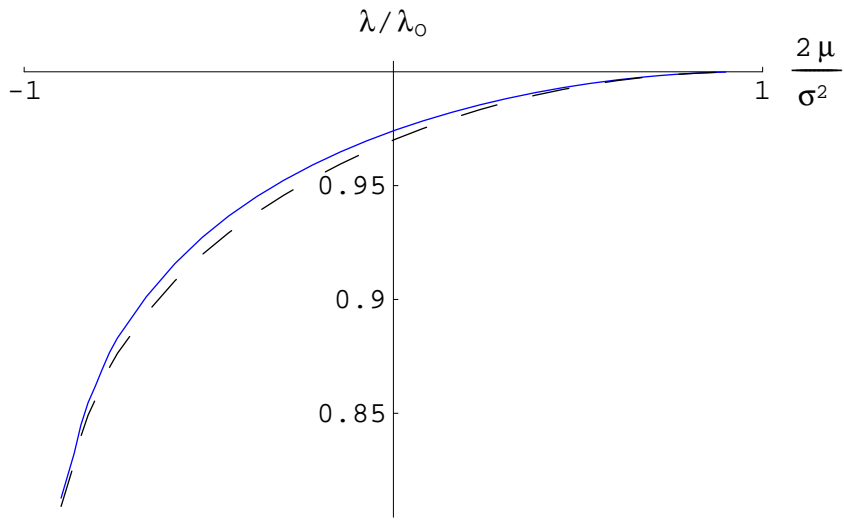}}
  \protect\caption{Comparison between the capital growth rate of the
  approximate solutions $\lambda$ rescaled with the exact one $\lambda_O$
  as a function of $2 \mu/ \sigma^2$ for $\gamma =0.1$.
  The dashed line indicates the {\it absolute} value case 
  (see appendix {\bf B}) and 
  the full line the {\it relative} value case.}
  \label{fig:lyap comparisons}
 \end{center}
\end{figure} 
 
In Figure~\ref{fig:lyap comparisons} we
compare the optimal capital growth rate 
with the approximate ones
obtained by the
two approaches ({\it absolute} and {\it relative}).
We notice that the differences between the two
are negligible
even for a so large (and unrealistic) transaction cost.
Furthermore the error committed 
considering these approximate solutions 
instead of the exact one is small
even for large $\gamma$.
As we would expect,
it is then reasonable, except for very small values of $\mu$, 
to limit a trading rule to a time independent $l$,
if no memory is present in the return process.

\section{Conclusions}

We have considered in this paper a new point of view
to treat the transaction costs problem.
This approach focuses only on the times the trader modifies
his position.
In the diversification of his portfolio the asset value (of all assets!)
depends on the trade (bid or ask) he is performing and
then it is {\it relative} to the market operator.
This approach presents several advantages 
compared with the ``traditional'' convention of the
{\it absolute} value.
 
The portfolio selection has been transformed to  
a Markovian problem in absence of costs.
We have connected the quantities of interest
(growth rate and optimal portfolio strategy)
to the average exit times and probabilities  
to reach a barrier.
If the returns can be modeled by a Wiener process,
as assumed in this paper,
both quantities 
can be computed using elementary
probability theory,  
allowing to have analytic formulas 
for both the
strategy ($l$ and $\Delta$s) 
and capital growth rate of the optimal solution.
More generally, however, 
the same technique can be easily extended to all the
utilities in the HARA class and
this approach allows a straightforward generalization
to the case with memory;
it is then particularly relevant in the case of a real market,
where exit times and probabilities to
reach a barrier can be measured directly 
from a historical dataset~\cite{BavVer}. 
 
We have shown that the exact solution of the problem 
in presence of transaction costs
breaks the time invariance of the investment.
However such a strategy is not feasible in practice
because the speculator should modify
his portfolio after a time which is on average zero.
In subsection {\bf 4.2}  
we have considered a suboptimal strategy 
feasible for a speculator who wants to use it.
This strategy, 
where the broken symmetry is restored,
shows small differences in the capital growth rate
with respect to the optimal one and
involves only 
a finite number of transactions in finite time.

We want to stress here that in real markets the 
variation of the returns are discrete and 
of the same order of magnitude
as the transaction cost $\gamma$.
To extend
this approach to this more realistic
case is enough to consider
the probability and the average time to exit from 
the barriers
in the discrete case shown in appendix {\bf A}.

Even in the case the costs to transact
are included,
the selection of a log-optimal portfolio  
depends on 
the information {\it \`a la} Shannon of the asset price:
in the general case this quantity can be measured on real data
and it can be computed analytically if an elementary 
model for the returns is assumed.
The Wiener process 
can be considered a toy model for asset returns,
nevertheless for a speculator who waits until 
relevant changes in the returns appear,
this model catches the essential features
of the portfolio selection,
leading to a deeper understanding
of the role of approximate but feasible 
strategies. 

\section*{Acknowledgment}

We thank Markus Abel for a careful reading of the manuscript,
Massimo Cencini for an enlightening conversation about Markovian processes,
Filippo Cesi for a nice discussion on exit times in Wiener processes,
Michele Pasquini, Maurizio Serva and Davide Vergni
for several comments and suggestions on the subject,
Angelo Vulpiani for a huge number of discussions about Kelly's
strategies and my father for a never-ending encouragement during
all this work.

\section*{Appendix A}
\appendix

In this appendix we compute the probabilities
and the average exit times of a Wiener process 
as limit of a random walk on a one dimensional lattice;
in this case 
both quantities can be obtained with elementary probability theory
(see for example Cap. 14 of~\cite{Feller}).

The walker 
goes after a time step $\epsilon$ 
to the right of a lattice step $\sigma \sqrt{\epsilon}$ with probability 
\[
p_\epsilon = \frac{1}{2} \left( 1 + \frac{\mu}{\sigma} \sqrt{\epsilon}\right)
\]
and to the left with probability $1-p_\epsilon$, where $\mu$ and $\sigma$
are the parameters of the Wiener process~(\ref{eq:wiener}).

Starting from zero 
the probability
to reach a barrier
situated at $-\Delta^-$ 
before hitting a barrier at $\Delta^+$ is
\[
 \pi_\epsilon(\Delta^+,\Delta^-) = 
  \left \{ {\begin {array} {ccc}
    {\displaystyle \frac{\rho_\epsilon^{\Delta^+}-1}{\rho_\epsilon^{\Delta^++\Delta^-}-1} }
	& {\mathrm if} & \mu \neq 0 \\
    {\displaystyle \frac{\Delta^+}{\Delta^++\Delta^-} }
	& {\mathrm if} & \mu = 0 
		\end{array} } \right.
\]
where
\[
\rho_\epsilon=\left(
\frac{p_\epsilon}{1-p_\epsilon} \right)^{\frac{1}{\sigma \sqrt{\epsilon}}} \,\,.
\]

The average time to exit from one of the two barriers is
\[
 \tau_\epsilon(\Delta^+,\Delta^-) = 
  \left \{ {\begin {array} {lcl}
   {\displaystyle \frac{1}{\mu}} \left[\Delta^+-(\Delta^++\Delta^-)\pi_\epsilon \right] & {\mathrm if} & \mu \neq 0 \\
   {\displaystyle \frac{1}{\sigma^2}} 
	\Delta^+ \Delta^- & {\mathrm if} & \mu = 0 \,\,.
		\end{array} } \right. 
\] 

Performing the limit $\epsilon \to 0$ one recovers the wiener process
defined in~(\ref{eq:wiener}) and defining
\[
\rho \equiv  \lim_{\epsilon \to 0} \rho_\epsilon = \exp \left[ \frac{2 \mu}{\sigma^2} \right] \,\,.
\]
we obtain the quantities of interest.
The probability to hit the lower barrier is 
\begin{equation}
 \pi(\Delta^+,\Delta^-) = 
  \left \{ {\begin {array} {ccl}
    {\displaystyle \frac{\rho^{\Delta^+}-1}{\rho^{\Delta^++\Delta^-}-1} }
	& {\mathrm if} & \mu \neq 0 \\
    {\displaystyle \frac{\Delta^+}{\Delta^++\Delta^-} }
	& {\mathrm if} & \mu = 0 \,\,.
		\end{array} } \right. \,\,
   \label{eq:pi}
\end{equation}
and the average exit time is
\begin{equation}
 \tau(\Delta^+,\Delta^-) = 
  \left \{ {\begin {array} {lcl}
   {\displaystyle \frac{1}{\mu}} \left[\Delta^+-(\Delta^++\Delta^-)\pi \right] & {\mathrm if} & \mu \neq 0 \\
   {\displaystyle \frac{1}{\sigma^2}} \Delta^+ \Delta^- & {\mathrm if} & \mu = 0 \,\,.
		\end{array} } \right. 
   \label{eq:tau}
\end{equation}
We observe that both the probability~(\ref{eq:pi}) 
and the exit time~(\ref{eq:tau}) 
are continuous in $\mu$.

We define martingale probability 
\begin{equation}
   q(\Delta^+,\Delta^-) = 
	{\displaystyle \frac{1-e^{-\Delta^+}}{1-e^{-\Delta^+-\Delta^-}}} \,\,,
\label{eq:martingale}
\end{equation} 
the one with respect to which the exponentiated return is a martingale process,
i.e.
\[
   q e^{-\Delta^-} +(1-q) e^{\Delta^+}=1\,\,.
\]

\section*{Appendix B}
\appendix

In this appendix we consider the case in which
the costs are due to the difference between 
the {\it absolute} asset value $S_t$ 
and the price the trader finds in the market
when he sells ($S^b_t$) or buys ($S^a_t$) assets.
The main advantage of this approach is that one can consider 
the time evolution of a unique ``true price'' $S_t$, 
and the trader diversifies his portfolio according to it 
taking into account the costs
every time he modifies his position.
The costs faced for each asset traded are
\[	
\begin {array} {ccl}
	(e^{\gamma_a} -1) S_{k+1} & {\mathrm if} & \eta=a\\
	(1 - e^{-\gamma_b}) S_{k+1} & {\mathrm if} & \eta=b \,\, .
\end{array}   
\]

After having payed the costs
the capital at the next $\Delta$-trading time will be worth:
\begin{equation}	
W_{k+1} = \left\{ \begin {array} {lcl}
	\left[ 1 + l_k (e^{-\Delta_k^-}-1) \right] W_k -
		     (e^{\gamma_a}-1) \left[ l_{k+1} W_{k+1} - 
		       l_k e^{-\Delta_k^-} W_k \right] 
	& {\mathrm if} & \eta = a\\
	\left[ 1 + l_k (e^{\Delta_k^+}-1) \right] W_k -
		     (1-e^{-\gamma_b}) \left[ 
		       l_k e^{\Delta_k^+} W_k - l_{k+1} W_{k+1} \right]
	& {\mathrm if} & \eta = b \,\,
 \end{array}  \right. 
\label{eq:absolute process}
\end{equation}
i.e. the capital evolves as in absence of 
costs~(\ref{eq:process no costs}),
but now the trader pays the costs  
on the assets he has bought or sold.

We can rewrite equation~(\ref{eq:absolute process}) in a multiplicative
form as in~(\ref{eq:process no costs}):
\begin{equation}	
W_{k+1} = \left\{ \begin {array} {ccl}
		{\displaystyle \frac { 1 - l_k + l_k 
			e^{\gamma_a-\Delta_k^-} }
		{1 - l_{k+1} + l_{k+1} e^{\gamma_a}}} \; W_k 
		& {\mathrm if} & \eta=a \\ 
		{\displaystyle \frac { 1 - l_k + l_k 
			e^{-\gamma_b + \Delta_k^+} }
		{1- l_{k+1} + l_{k+1} e^{-\gamma_b} } } \; W_k 
		& {\mathrm if} & \eta=b  \,\, .
 \end{array}  \right. 
\label{eq:absolute process II}
\end{equation}

To show this well known approach in a simple case,
we consider the ansatz of $l$ and $\{\Delta^-,\Delta^+\}$
to be time independent,
as in the approximate solution of the {\it relative} value approach.
The mean lyapunov exponent of the capital is
\begin{equation}
 h(l;\Delta^+,\Delta^-) =  p \ln { \displaystyle\frac { 1 - l + l e^{\gamma_a-\Delta^-}}
		{1 - l + l e^{\gamma_a}} } + 
    (1-p) \ln { \displaystyle\frac { 1 - l + l e^{-\gamma_b+\Delta^+}}
		{1 - l + l e^{-\gamma_b}} } \,\,
   \label{eq:lyapunov absolute costs}
\end{equation} 
where $p=\pi(\Delta^+,\Delta^-)$ 
is the probability~(\ref{eq:pi}) to exit from the lower barrier.

The optimal value of $l$ satisfies the equation:
\[
 \begin {array} {l}
0= p e^{\gamma_a}(e^{-\Delta^-}-1) 
\left[ 1+ l (e^{\Delta^+-\gamma_b}-1)\right]
\left[ 1+ l (e^{-\gamma_b}-1)\right]+ \\
\;\; (1- p) e^{-\gamma_b}(e^{\Delta^+}-1) 
\left[ 1+ l (e^{-\Delta^-+\gamma_a}-1)\right]
\left[ 1+ l (e^{\gamma_a}-1)\right]
\equiv d l^2 + f l + g \,\,. 
 \end {array}
\]

Solving this second order equation one obtains
\begin{equation}
 l(\Delta^+,\Delta^-) = \frac{-f+\sqrt{f^2 - 4 d g}}{2 d}\,\, ,
\label{eq:l ottimale}
\end{equation} 
which corresponds to the absence of costs solution in the limit
$\gamma \to 0$.
The other solution for $l$ 
lies outside the interval of values allowed
by a self-financing strategy. 

\begin{figure}[htb]
 \begin{center}
  \resizebox{0.7\textwidth}{!}{\includegraphics{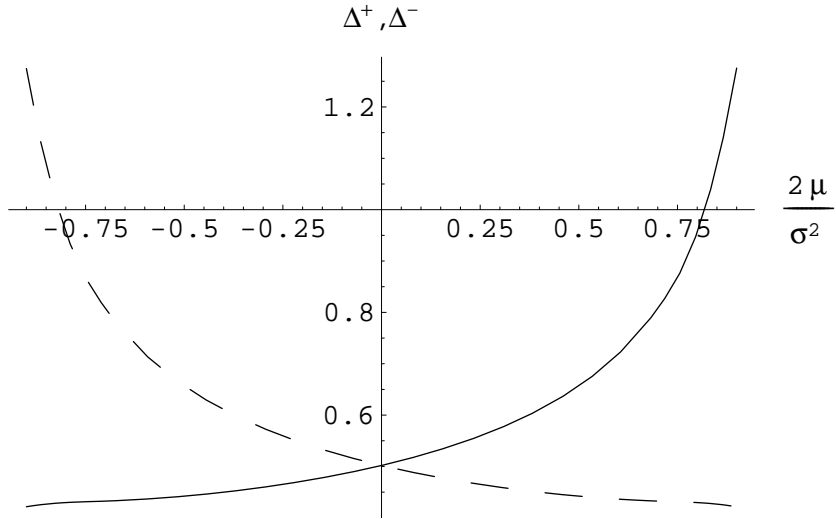}}
  \protect\caption{Optimal values of $\Delta^-$ (full line) and $\Delta^+$ (dashed line) {\it vs} $2 \mu/ \sigma^2$ for $\gamma = 0.01$ and
$\gamma_a=\gamma_b=\gamma/2$.}
  \label{fig:delta absolute costs}
\end{center}
\end{figure}

In Figure~\ref{fig:delta absolute costs} 
we show the optimal values for the barriers 
obtained maximizing numerically  the capital growth rate.
We notice that 
$(\Delta^-(-\mu),\Delta^+(-\mu))=(\Delta^+(\mu),\Delta^-(\mu))$ 
and $l(-\mu)=1-l(\mu)$, 
because, for $\gamma_a=\gamma_b=\gamma/2$, 
the growth rate~(\ref{eq:lyapunov absolute costs})
has the same symmetry in $\mu$  (see equation~(\ref{eq:simmetry})) 
of the {\it relative} approach.


\begin{thebibliography}{99}

\bibitem{TakKlaAss} M.~Taksar, M.~J.~Klass and D.~Assaf,
{\small\it A diffusion model for optimal portfolio
selection in presence of brokerage fees},
{\small\it Math. of Oper. Research} {\small\bf 13} (1988) 277--294. 

\bibitem{Constantinides} G.~M.~Constantinides,
{\small\it Multiperiod consumption and investment
behavior with convex transactions costs},
{\small\it Management Science} {\small\bf 25} (1991) 1127--1137.

\bibitem{DumLuc} B.~Dumas and E.~Luciano,
{\small\it An exact Solution to a Dynamic Portfolio Choice Problem
under Transaction Costs},
{\small\it J. of Finance} {\small\bf 46} (1991) 577--595.

\bibitem{AkiMenSul} M.~Akian, J.~L.~Menaldi and A.~Sulem,
{\small\it On an investment-consumption model with transaction costs},
{\small\it SIAM J. Control and Optim.} {\small\bf 34} (1996) 329--364.

\bibitem{Serva} M.~Serva,
{\small\it Optimal lag in dynamical investment},
{\small\it International J. of Theoretical and Applied Finance},
{\small\it to appear}.

\bibitem{Harrison} J.~M.~Harrison, 
{\small\bf Brownian motion and Stochastic Flow System}, 
(John Wiley \& Sons, Somerset N. J., 1991).

\bibitem{Karatzas} I.~Karatzas and S.~E.~Shreve 
{\small\bf Brownian motion and Stochastic Calculus}, 
(Springer--Verlag, 1985).

\bibitem{Kelly} Jr.~J.~L.~Kelly,
{\small\it A new interpretation of the Information Rate},
{\small\it Bell Syst. Tech. J.} {\small\bf 35} (1956) 917--926.

\bibitem{Shannon} C.~E.~Shannon,
{\small\it The mathematical theory of communication},
{\small\it Bell Syst. Techn. Journ.} {\small\bf 27} (1948) 379--423, 623--656.

\bibitem{BavPasSerVerVul}
R.~Baviera, M.~Pasquini, M.~Serva, D.~Vergni and A.~Vulpiani, 
{\small\it Efficiency in foreign exchange markets},
{\small\it Submitted to J. Financial Economics}, 
available in http://papers.ssrn.com (1999).

\bibitem{BadPol} R. Badii and A. Politi,
{\small\bf Complexity: hierarchical structures and scaling in physics}, 
(Cambridge University Press, Cambridge, 1997).

\bibitem{BavVerVul} R.~Baviera, D.~Vergni and A.~Vulpiani,
{\small\it Markovian approximation in foreign exchange markets},
{\small\it Submitted to J. International Money and Finance},
available in xlanl archives cond-mat/9903144 (1999).

\bibitem{Merton} R.~C.~Merton, 
{\small\it Optimum consumption and portfolio rules in a continuous-time model},
{\small\it J. of Econ. Theory} {\small\bf 3} (1971) 373--413.

\bibitem{Kullback} S.~Kullback and R.~A.~Leibler.
{\small\it On information and sufficiency},
{\small\it Ann. Math. Statist.} {\small\bf 27} (1956) 79--86.

\bibitem{BavVer} R.~Baviera,
{\small\it Weak efficiency in absence and in presence of transaction costs},
{\small\it in preparation}.

\bibitem{Kolmogorov} A.~N.~Kolmogorov, 
{\small\it On the Shannon theory of information transmission in the case
of continuous signals},
{\small\it IRE Trans. on Inform. Theory} {\small\bf IT-2} (1956) 102--108.

\bibitem{Feller} W.~Feller,
{\small\bf An introduction to probability theory and its applications I}, 
(John Wiley \& Sons, New York, 1971). 


\end{thebibliography}
\end{document}